\documentclass[10pt,twocolumn,twoside]{IEEEtran}

\usepackage[mathcal]{euscript}
\usepackage{amssymb}
\usepackage{amsfonts}
\usepackage{amsmath}
\usepackage{cite,url}
\usepackage[dvips]{graphicx}
\usepackage{color}
\usepackage{amsopn}
\usepackage{comment}

\begin{document}

\newtheorem{definition}{\it Definition}
\newtheorem{theorem}{\bf Theorem}
\newtheorem{lemma}{\it Lemma}
\newtheorem{corollary}{\it Corollary}
\newtheorem{remark}{\it Remark}
\newtheorem{example}{\it Example}
\newtheorem{case}{\bf Case Study}
\newtheorem{assumption}{\it Assumption}
\newtheorem{property}{\it Property}
\newtheorem{proposition}{\it Proposition}

\newcommand{\hP}[1]{{\boldsymbol h}_{{#1}{\bullet}}}
\newcommand{\hS}[1]{{\boldsymbol h}_{{\bullet}{#1}}}

\newcommand{\ba}{\boldsymbol{a}}
\newcommand{\baq}{\overline{q}}
\newcommand{\bA}{\boldsymbol{A}}
\newcommand{\bcA}{\boldsymbol{\cal A}}
\newcommand{\bb}{\boldsymbol{b}}
\newcommand{\bB}{\boldsymbol{B}}
\newcommand{\bc}{\boldsymbol{c}}
\newcommand{\bC}{\boldsymbol{C}}
\newcommand{\bcC}{\boldsymbol{\cal C}}
\newcommand{\bcE}{\boldsymbol{\cal E}}
\newcommand{\bcO}{\boldsymbol{\cal O}}
\newcommand{\bd}{\boldsymbol{d}}
\newcommand{\be}{\boldsymbol{e}}
\newcommand{\bh}{\boldsymbol{h}}
\newcommand{\bH}{\boldsymbol{H}}
\newcommand{\bl}{\boldsymbol{l}}
\newcommand{\bL}{\boldsymbol{L}}
\newcommand{\bm}{\boldsymbol{m}}
\newcommand{\bn}{\boldsymbol{n}}
\newcommand{\bo}{\boldsymbol{o}}
\newcommand{\bO}{\boldsymbol{O}}
\newcommand{\bp}{\boldsymbol{p}}
\newcommand{\bq}{\boldsymbol{q}}
\newcommand{\br}{\boldsymbol{r}}
\newcommand{\bR}{\boldsymbol{R}}
\newcommand{\bs}{\boldsymbol{s}}
\newcommand{\bS}{\boldsymbol{S}}
\newcommand{\bT}{\boldsymbol{T}}
\newcommand{\bu}{\boldsymbol{u}}
\newcommand{\bw}{\boldsymbol{w}}
\newcommand{\bcY}{\boldsymbol{\cal Y}}
\newcommand{\bx}{\boldsymbol{x}}
\newcommand{\by}{\boldsymbol{y}}

\newcommand{\balpha}{\boldsymbol{\alpha}}
\newcommand{\bbeta}{\boldsymbol{\beta}}
\newcommand{\blambda}{\boldsymbol{\lambda}}
\newcommand{\bLambda}{\boldsymbol{\Lambda}}
\newcommand{\bOmega}{\boldsymbol{\Omega}}
\newcommand{\bTheta}{\boldsymbol{\Theta}}
\newcommand{\bphi}{\boldsymbol{\phi}}
\newcommand{\btheta}{\boldsymbol{\theta}}
\newcommand{\btau}{\boldsymbol{\tau}}
\newcommand{\bvarpi}{\boldsymbol{\varpi}}
\newcommand{\bvarphi}{\boldsymbol{\varphi}}
\newcommand{\bpi}{\boldsymbol{\pi}}
\newcommand{\bpsi}{\boldsymbol{\psi}}
\newcommand{\brho}{\boldsymbol{\rho}}
\newcommand{\bxi}{\boldsymbol{\xi}}

\newcommand{\cA}{{\cal A}}
\newcommand{\cB}{{\cal B}}
\newcommand{\cC}{{\cal C}}
\newcommand{\cD}{{\cal D}}
\newcommand{\cE}{{\cal E}}
\newcommand{\cF}{{\cal F}}
\newcommand{\cG}{{\cal G}}
\newcommand{\cH}{{\cal H}}
\newcommand{\bcH}{\boldsymbol {\cal H}}
\newcommand{\cK}{{\cal K}}
\newcommand{\cO}{{\cal O}}
\newcommand{\cR}{{\cal R}}
\newcommand{\cS}{{\cal S}}
\newcommand{\dcS}{\ddot{{\cal S}}}
\newcommand{\ds}{\ddot{{s}}}
\newcommand{\cT}{{\cal T}}
\newcommand{\cU}{{\cal U}}
\newcommand{\cW}{{\cal W}}
\newcommand{\cY}{{\cal Y}}
\newcommand{\wt}[1]{\widetilde{#1}}

\newcommand{\mA}{\mathbb{A}}
\newcommand{\mE}{\mathbb{E}}
\newcommand{\mG}{\mathbb{G}}
\newcommand{\mR}{\mathbb{R}}
\newcommand{\mS}{\mathbb{S}}
\newcommand{\mU}{\mathbb{U}}
\newcommand{\mV}{\mathbb{V}}
\newcommand{\mW}{\mathbb{W}}

\newcommand{\uq}{\underline{q}}
\newcommand{\ubq}{\underline{\boldsymbol q}}

\newcommand{\red}[1]{\textcolor[rgb]{1,0,0}{#1}}
\newcommand{\gre}[1]{\textcolor[rgb]{0,1,0}{#1}}
\newcommand{\blu}[1]{\textcolor[rgb]{0,0,1}{#1}}


\title{Distributed Optimization for Energy-efficient Fog Computing in the Tactile Internet}

\author{Yong~Xiao, \IEEEmembership{Senior Member, IEEE} and Marwan Krunz, \IEEEmembership{Fellow, IEEE}

\thanks{Manuscript received April 15, 2018; accepted August 18, 2018.  Part of the results in this paper was presented at the IEEE INFOCOM Conference, Atlanta, May 2017\cite{XY2017InfocomFogComp}.
This research was supported in part by NSF (grants IIP-1265960, CNS-1409172, CNS-1513649, CNS-1563655, and CNS-1731164) and by the Broadband Wireless Access \& Applications Center. Any opinions, findings, conclusions, or recommendations expressed in this paper are those of the author(s) and do not necessarily reflect the views of NSF.

Y. Xiao is with the School of Electronic Information
and Communications at the Huazhong University of Science and Technology, Wuhan, China (e-mail: yongxiao@hust.edu.cn).

M. Krunz is with the Department of Electrical and Computer Engineering at the University of Arizona, Tucson, AZ (e-mail: krunz@email.arizona.edu) and is also affiliated with the
University Technology Sydney (UTS), Sydney, Australia} 
}
\maketitle
\begin{abstract}
Tactile Internet is an emerging concept that focuses on supporting high-fidelity, ultra-responsive, and widely available human-to-machine interactions. To reduce the transmission latency and alleviate Internet congestion, fog computing has been advocated as an important component of the Tactile Internet. 
In this paper, we focus on energy-efficient design of fog computing networks that support low-latency Tactile Internet applications. We investigate two performance metrics: {\em Service response time of end-users} and {\em power usage efficiency of fog nodes}. We quantify the fundamental tradeoff between these two metrics and then extend our analysis to fog computing networks involving cooperation between fog nodes.  
%
%
%
We introduce a novel cooperative fog computing concept, referred to as {\em offload forwarding}, 
in which a set of fog nodes with different computing and energy resources can cooperate with each other. The objective of this cooperation is to balance the workload processed by different fog nodes, 
further reduce the service response time, and improve the efficiency of power usage. 
%
We develop a distributed optimization framework based on {\em dual decomposition} to achieve the optimal tradeoff. Our framework does not require fog nodes to disclose their private information nor conduct back-and-forth negotiations with each other. Two distributed optimization algorithms are proposed. One is based on the subgradient method with dual decomposition and the other is based on distributed alternating direction method of multipliers via variable splitting (ADMM-VS). We prove that both algorithms can achieve the optimal workload allocation that minimizes the response time under the given power efficiency constraints of fog nodes.
%
Finally, to evaluate the performance of our proposed concept, we simulate a possible implementation of a city-wide self-driving bus system supported by fog computing in the city of Dublin. The fog computing network topology is set based on a real cellular network infrastructure involving 200 base stations deployed by a major cellular operator in Ireland. Numerical results show that our proposed framework can balance the power usage efficiency among fog nodes and 
reduce the service latency for users by around 50\% in urban scenarios.

\end{abstract}
\vspace{-0.2in}

\section{Introduction}
\label{Section_Introduction}
With the widespread deployment of high-performance computing infrastructure and advancement of networking and communication technology, it is believed that the vision of the Tactile Internet (TI) will soon become a reality, transforming the existing content-delivery-based Internet into skill-set delivery-based networks\cite{Simsek2016TactileInternet}. According to the Next-Generation Mobile Network (NGMN) Alliance\cite{NGMN5GWhitePaper}, TI is defined as the capability of remotely delivering real-time control over both real and virtual objects as well as physical haptic experiences through the Internet. 
It will be able to contribute to the solution of many complex challenges faced by our society, enabling novel services and applications that cannot fit well in the current state-of-the-art networking and cloud computing architectures. Examples of these applications include long-distance education with immersive learning experience, high-precision remote medical diagnosis and treatment, high-sensitive industry control and automation, collision-avoidance for high-speed autonomous vehicles, high-fidelity virtual/augment reality (VR/AR), etc.   
Recent analysis shows that the TI has the potential to generate up to \$20 trillions global market, accounting for around 20\% of the global GDP\cite{Maier2016TactileInternet}.
Because the TI will provide critical services, it needs to be extremely reliable, ultra-responsive, and widely available. More specifically, according to International Telecommunication Union (ITU), the TI must support latencies as low as 1 ms, the reliability of around one second of outage per year, enhanced security, as well as sufficient computational resources within the range of a few kilometers from each end-user\cite{ITU2014TactileInternet}. 
%

Fog computing has been recently introduced as a promising solution to accommodate the stringent requirements of the TI. 
It complements the cloud infrastructure by involving a large number of low-cost, often decentralized devices, commonly referred to as fog nodes, to perform computing, storage, control, and network functions closer to end-users\cite{Chiang2016FogCompu,Dastjerdi2016FogComp,Yi2015Fog2}. Fog nodes embody a variety of devices between end-users and cloud data centers (CDCs), including routers, smart gateways, access points (APs), base stations (BSs), as well as portable devices such as drones, robots, and vehicles with computing and storage capabilities\cite{Dinh2013survey}. 
The success of the TI will hinge on 
widespread deployment of fog nodes with high computational capabilities and reliable energy supply \cite{NGMN5GWhitePaper}. 
However, the theoretical foundations for optimizing distributed fog computing systems to meet the demands of the TI are still lacking. 
In particular, 
computationally intensive services requiring low latencies generally demand more energy consumption from fog nodes. At the same time, many TI applications involve portable devices such as robots, drones, and vehicles with limited power supplies.  
%
%
%
In addition, the fast growing power consumption of information and communication technologies  and its impact on climate change have recently raised significant concerns in both industry and academia\cite{ETSI2014MEC, Vaquero2014FogComp}. Existing cloud data centers in the US have already constituted more than 2\% of the country's total electricity usage. Power consumption is expected to be significantly increased with the deployment of a large number of fog computing servers throughout the world. How to improve the efficiency of the power usage for fog computing networks while taking into consideration the stringent requirements of the TI services  is still an open problem.

Another issue is that, 
in contrast to other applications, 
workload generated by the TI can exhibit much higher temporal and geographical variations due to the bursty nature of human-generated traffic. For example, an autonomous vehicle that is trying to pass another will create a much larger computational workload and require much lower latency service compared to other vehicles that stick to pre-planned routes. 
How to efficiently distribute and orchestrate the workload of different fog nodes for parallel execution under real-time constraints is still an open issue. 

In this paper, we take steps towards addressing the above issues. In particular, we study energy-efficient workload offloading for fog computing systems that support the TI applications and services. 
We focus on optimizing two important performance metrics: (1) {\em Service response time}, including the round-trip transmission latency between users and fog nodes, queueing delays, and workload transmission and forwarding latency among fog nodes as well as that between fog nodes and CDCs; and (2) 
{\em fog nodes' power efficiency}, measured by the amount of power consumed by fog nodes to process a unit of workload. 
We perform detailed analysis under different scenarios and derive the optimal amount of workload to be processed by fog nodes so as to minimize the response time under a given power efficiency. 
We quantify the fundamental tradeoff between these two metrics. 

To address the issue of skewed workload distribution among fog nodes, we study a cooperative setting in which 
the workload of a fog node can be partially processed by other nodes in proximity. We observe that the response time and power-efficiency tradeoff is closely related to the cooperation strategy among fog nodes.  
Accordingly, we propose a novel cooperation strategy called {\em offload forwarding}, in which each fog node can forward a part or all of its unprocessed workload to other nearby fog nodes, instead of always forwarding workload that exceeds its processing capability to a remote CDC. We study the offload allocation problem in which all fog nodes jointly determine the optimal amount of workload to be forwarded and processed by each other to further reduce the response time. 

Based on our analysis, we observe that, for most TI applications, it is generally impossible to optimize the workload distribution among fog nodes in a centralized fashion due to the following reasons: (1) Deploying a central controller to calculate the amounts of workload processed by every fog node 
may result in intolerably high information collection and coordination delay as well as high computation complexity at the controller; (2) the workload received by each fog node can be highly dynamic, and constantly exchanging information about workload as well as computational resource availability among fog nodes can result in network congestion; and (3) fog nodes may not want to reveal their private information. 
Motivated by these observations, we propose a novel distributed optimization framework for cooperative fog computing based on {\em dual decomposition}. Our proposed framework does not require fog nodes to have back-and-forth negotiation or disclose their private information. 
Two distributed algorithms are developed under the proposed framework. The first one is based on the {\em subgradient method with dual decomposition}. We show that this algorithm converges to the globally optimal solution with low computational complexity at each fog node. 
To further improve the convergence rate, we propose another distributed algorithm based on {\em distributed alternating direction method of multipliers via variable splitting} (ADMM-VS). We prove that 
ADMM-VS can converge to the globally optimal solution in linear time with a slight increase in the computational complexity at each fog node. 


Motivated by the fact that the self-driving vehicle has been considered as one of the key use cases for the TI\cite{ITU2014TactileInternet, Simsek2016TactileInternet}, as a case study, we evaluate the performance of our framework by simulating a city-wide implementation of a self-driving bus system supported by a fog computing network. 
We analyze over 2500 traffic images of 8 existing bus routes operated at the city of Dublin and consider the scenario that these traffic data can be submitted and processed by a fog computing network deployed in a real wireless network infrastructure consisting of over 200 base stations of a major cellular operator in Ireland to ensure safe and efficient decision making and driving guidance for all the buses. 
We evaluate the service response time and power efficiency of fog computing networks in different areas of the city with different densities of fog node deployment.
Numerical results show that our algorithms can almost double the workload processing capacity of fog nodes in urban areas with high density of fog node deployment. To the best of our knowledge, this is the first work that studies distributed workload allocation among cooperative fog nodes with energy efficiency awareness. 

\vspace{-0.1in}
\section{Related Work}
\label{Section_RelatedWorks}
ITU and NGMN Alliance identify fog computing as one of the key components for the TI to achieve ultra-low service latency for users\cite{Simsek2016TactileInternet, ITU2014TactileInternet,Maier2016TactileInternet}. In contrast to CDCs, fog nodes can be built much closer to users so the workload transmission latency can be significantly reduced. However, due to the limited computational capability of each fog node, offloading a large amount of workload to fog nodes will result in a high processing delay. 
Therefore, previous works focused on how to optimize resource provisioning of fog nodes so as to reduce the processing delay. For example, in \cite{Satyanarayanan209Cloudlet}, a virtual machine (VM) synthesis approach was proposed to allow each end user to quickly provision the resources of neighboring fog nodes and create the required VM images to support the requested service. The resource provisioning problem for a CDC network has been modeled in \cite{Zheng2015BidCloud} as an auction-based market, where users develop bidding strategies to compete for the CDC 
at low costs. In \cite{Aazam2015Fog}, a service-oriented resource estimation and management framework for fog computing was introduced to maximize the resource utilization of the CDCs.
The authors in \cite{Cuervo2010CompOffload} introduced a system that allows fine-grained energy-aware offloading of users' mobile codes to the infrastructure. In \cite{Lewis2014CompOffloading}, resource provisioning was investigated for tactical cloudlets. A provisioning mechanism was proposed for the infrastructure to support computation offloading and data staging at the tactical edge. In \cite{Tong2016}, a hierarchical architecture was proposed in which edge cloud servers are organized into different tiers according to their distances to the edge. If the workload received by an edge cloud server of a given tier exceeds its computational capacity, the extra workload is forwarded to higher-tier servers. In \cite{Tan2017MobileEdge}, a scalable online algorithm was introduced to optimize the workload dispatched to a  fog node.

One of the premises of fog computing 
is that a large number of fog nodes are widely deployed across a large geographical area. This, however, can result in a significant increase in energy consumption. In addition, some fog computing systems such as computational ferry\cite{Monfared2015ComputFerry}, mobile cloudlet\cite{Zhang2015MobileCloudlet} and fog nodes deployed in tactical environments\cite{Echeverria2015TacticCloudlet}, have limited access to reliable power sources (e.g., power grid). Therefore, developing energy-efficient solutions for fog computing is critical to ensure the sustainability, availability, and ubiquity of services. In \cite{Igder2016EnergyFog}, an energy adaptive scheme was proposed for fog nodes to operate at different transmit powers with variable data rates. The authors in \cite{Jalali2016EnergyFog} compared the energy consumption of applications using centralized CDCs with applications that use nano data centers under the fog computing architecture. They verified that the most energy-efficient strategy for content storage and distribution in cloud applications is a combination of centralized data centers and distributed fog nodes.
In this paper, we evaluate our proposed fog node cooperation framework by simulating a fog computing-supported self-driving bus system.
Most existing self-driving vehicle platform, including Google's Waymo self-driving car project\cite{Google2016Waymo} and Stanford's Junior self-driving vehicle platform\cite{Teichman2011SelfDrivingCar}, focus on the scenarios that the driving decision will be made by 
an on-board computer according to a pre-set policy and/or human instructions\footnote{Society of Automotive Engineers (SAE) International defines autonomous driving systems into six levels: no automation, driver assistance, partial automation, conditional automation, high automation, and full automation. Note that even when a vehicle has been defined as full automation, human intervention is required under emergency situations\cite{Ilkova2017SelfDrive}.}. Recently, fog computing-supported self-driving systems have attracted significant interest\cite{Yuan2018SelfDrivingCar, Grewe2017SelfDrivingCars}. In contrast to the existing on-vehicle computing solution, fog computing-assisted driving guidance and decision making can support high-performance image and video processing as well as high capacity storage for storing high-definition maps with instantaneous traffic updating.

So far, the impact of the energy-efficient fog computing design on the service response time has not been well investigated. Motivated by this observation, in this paper, we study the relationship between the response time of users and the power efficiency of fog nodes considering the possibility of fog node cooperation to support various TI applications. 

\vspace{-0.1in}
\section{Fog Computing-supported Tactile Internet Architecture}
\label{Section_SystemModel}

\begin{figure}
\centering
\includegraphics[width=3.5 in]{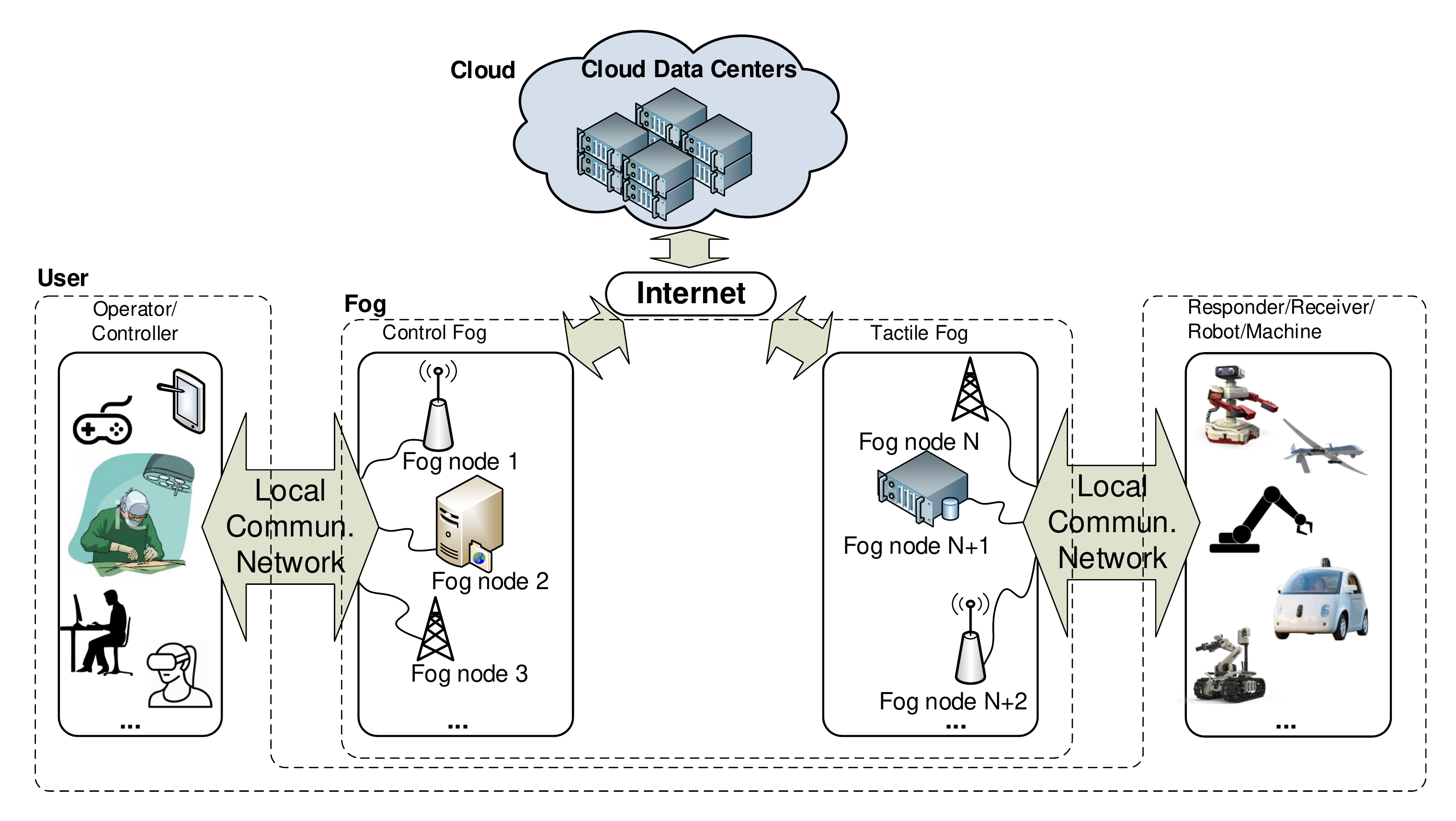}
\vspace{-0.3in}
\caption{Fog computing-supported Tactile Internet architecture.} 
\label{Figure_FogCompuArchitecture}\vspace{-0.2in}
\end{figure}

A generic fog computing-supported Tactile Internet architecture 
consisting of four major components as illustrated in Figure \ref{Figure_FogCompuArchitecture}\cite{Simsek2016TactileInternet, ITU2014TactileInternet,Maier2016TactileInternet}: 

\noindent
{\em 1) Operator}--A human operator and/or human-to-machine (H2M) interface that can manipulate virtual and/or real objectives using various input signals such as human gestures, touch, and voice. In some applications such as self-driving vehicles and autonomous robots, operators can also correspond to pre-calculated policies that can mimic the human behaviors/decision making processes. Operators can also expect feedback within a given time duration depending on the particular applications. 
    For example, a real-time AR/VR game may require as low as 1ms response time. Other applications, such as remotely controlled robots, can tolerate up to one second response time.

\noindent
{\em 2) Responder}--One or multiple teleoperators (remotely controlled robots, machines, drones, etc.) that can be directly controlled by the operators. Responders interact with the environment and send feedback signals to the operators.

\noindent
{\em 3) Fog}--The composition of a number of low-cost fog nodes that are characterized by limited computing capabilities and power supplies. Fog nodes are deployed closer to end-users. 
    According to the types of users served by fog nodes, the fog can be further divided into {\em control fog} and {\em tactile fog}. Control fog consists of fog nodes that can support computation-intensive services, 
    such as 
    analog-to-digital conversion, coding, signal processing, data compression, etc. Tactile fog consists of the fog nodes that are responsible for processing, compressing, and sending feedback data generated through the interactions between responders and the environment.
Fog nodes may take the form of mini-servers  
within the wireless edge network infrastructure, reside in base stations (BSs), roadside units (RSUs), etc. Each fog node serves a distinct set of users in its coverage area. Users first submit their workloads to the closest fog node. Each fog node will then need to carefully decide the amount of workload it can process locally. If multiple closely located fog nodes can communicate with each other 
(e.g., using high-speed backhaul connection links in a cellular network), some fog nodes may forward part of their workload to other nearby nodes to further improve the processing capability and balance the workload among nodes. 

\noindent
{\em 4) Cloud}--Large-scale CDCs equipped with powerful processing units. These data centers are often built in remote areas, far from end users.

Fog nodes and users (operators or responders) may correspond to the same type of devices. For example, in some distributed mobile computing grid systems (e.g., IBM's world community grid project), the idle computing resources of some computers (i.e., fog nodes) can be used by other computers (i.e., users)  to perform computationally intensive tasks. 

%
We consider a fog computing system that contains 
a set of $N$ fog nodes ${\cal F} = \{1, 2, \ldots, N\}$. 
Any user can be associated with one or more of these fog nodes. The association between users and fog nodes can be made based on physical proximity, channel conditions, 
or prior agreements between users and network service provider. For example, if the fog is deployed by the cloud provider, users can send their service requests to CDCs following the same procedure of a traditional cloud computing system. The cloud provider can then delegate one or more nearby fog nodes to process the workload submitted by these users. Each fog node $j$ can process a nonnegative portion $\alpha_j$ of its received workload using its local resources. Remaining workload, if any, is forwarded to the cloud. Note that $\alpha_j = 1$ means that fog node $j$ will process all its received workload. 
The workload arrival rate at each fog node $j$, denoted by $\lambda_j$, is assumed to be fixed. 

We focus on two performance metrics: 

\noindent
{\em 1) (Service) response time of end-users}: 
    The response time includes the round-trip time for transmitting the workload between a user and the associated fog node as well as the queueing delay at the fog. 
    Given their proximity to to users, fog nodes are likely to exhibit smaller transmission times than 
    remote CDCs. 
    However, due to their limited resources, 
    fog nodes that process a large amount of workload will likely have a long queueing delay. Therefore, it is important to balance the workload offloaded by fog nodes. Note that the response time associated with fog node $i$, denoted as $R_i \left( \alpha_i \right)$, depends on the portion of workload locally processed by fog node $i$. 

\noindent
{\em 2) Power efficiency of fog nodes}: 
We consider the power efficiency 
by the amount of power spent on processing a unit of received workload. 
Maximizing the power efficiency amounts to minimizing the power consumption for processing a given workload. It is known that the total amount of power consumed by any electronic device (e.g., a fog node) depends on the power usage effectiveness (PUE) as well as the static and dynamic power consumption. PUE is the input power from the power grid divided by the power consumption of the given device. Static power consumption, also called leakage power, is mainly caused by the leakage currents, and is unrelated to the usage of the computing resources at a fog node. Dynamic power consumption is the result of the circuit activity and is determined by the activity of computing resources. Let $e_i$ and $w^S_{i}$ be the PUE and static power consumption of fog node $i$, respectively. Let $w^D_i$ be the dynamic power consumed by fog node $i$ to offload each unit of workload. 
    We can write the total power consumption of fog node $i$ per time unit as $w_i = e_i \left( w^S_{i} + w^D_i \alpha_i \lambda_j \right)$.
%
    The power efficiency of fog node $i$ can then be written as
    \begin{eqnarray}
    \eta_i \left( \alpha_i \right) = {w_i \over \alpha_i \lambda_i} =  e_i \left( {w^S_{i} \over \alpha_i \lambda_i} + w^D_i \right).
    \label{eq_etaj}
    \end{eqnarray}
One of the main objective of this paper is to develop workload allocation strategies for 
determine the appropriate portion of workload to be processed locally so as to minimize the response time under given power-efficiency constraints. 
Formally, each fog node $i$ tries to find the optimal value $\alpha^*_{i}$ by solving the following optimization problem:
\begin{eqnarray}
\alpha^*_{i} &=& \arg \min\limits_{\alpha_{i} \in [0, 1]} R_{i} \left( \alpha_{i} \right)
\label{eq_OptimProb_ResponseTimeSingleNode} \\
&&\;\;\;\; \mbox{s.t.}\;\;  \eta_i \left( \alpha_i \right) \le \bar \eta_i, \nonumber 
\end{eqnarray}
where 
$\bar \eta_i$ is the maximum power efficiency that can be supported by the hardware of fog node $i$. In Section \ref{Section_1Fognode}, we will give a more detailed discussion of the response time of fog node $i$ under different scenarios.

As mentioned earlier, different fog nodes can have different workload arrival rates. Therefore, allowing fog nodes to cooperate with each other and jointly process their received workload can further improve the overall workload processing capability. Specifically, fog nodes that receive more workload than their processing capabilities can seek help from nearby fog nodes with surplus computing resources. 
The main objective in this case is to minimize the average response time of users associated with all cooperative fog nodes. 
The total amount of workload processed by each fog node in this case will not only depend on its own received workload, but also on the workload forwarded from other fog nodes. We can write the response time of fog node $i$ under cooperation as $R^C_{i} \left( \balpha \right)$ where $\balpha = \langle {\alpha_{1}}, \alpha_2 \ldots, \alpha_{N} \rangle$. The optimal workload distribution  under cooperative fog computing can then be written as
\begin{eqnarray}
&&\balpha^* = \arg \min\limits_{\balpha} \sum\limits_{i \in \cF}  R^C_{i} \left( \balpha \right)
    \label{eq_OptimProb_ResponseTimeMultiNode}\\
    &&\;\;\;\; \mbox{s.t.}\;  \eta_i \left( \alpha_i \right) \le \bar \eta_i, 0 \le \alpha_{i} \le 1,  \forall i \in {\cal F}. \nonumber
\end{eqnarray}
Later on, we provide a more detailed discussion of the strategies for cooperative fog computing. 

\vspace{-0.1in}
\section{Response Time and Power Efficiency Tradeoff} 
\label{Section_1Fognode}

\subsection{Response Time Analysis and Minimization}
Let $\tau^u_{j}$ be the average round trip time (RTT) between fog node $j$ and its users. Typically, fog nodes and CDCs have fixed locations. Thus, we assume the average workload transmission time between each fog node $j$ and the cloud can be regarded as a constant denoted as $\tau^f$. 
Node $j$ can directly forward its received workload to CDCs through the backbone IP network\cite{Chiang2016FogCompu}. In this case, the fog computing network becomes equivalent to the traditional cloud computing network with all the workload being processed by the cloud. As mentioned before, 
CDCs are generally installed with high-performance workload processing units, and therefore their processing times are 
much smaller than the workload transmission time\cite{Keller2014QDelay}. For simplicity, we ignore the processing time of CDCs.  
In this case, the response time of fog node $j$ can be written as 
${R^{W1}_{j}} = \tau^u_{j} + \tau^f$.
%
Since in this case fog node $j$ does not activate any computing resources to process its received workload, the power efficiency will not depend on the response time. 

In another extreme case, node $j$ may process all its received workload using its local computing resources, 
i.e., $\alpha_j = 1$.
If we follow a commonly adopted setting and consider an M/M/1 queueing system for each fog node to process the received request, 
we can write the response time of fog node $j$ as
${R^{W2}_{j}}\left( \lambda_{j} \right) = \tau^u_{j} + {1 \over {\mu_j -  \lambda_j}}$ where $\mu_j$ os the maximum amount of workload that can be processed by the on-board computing resources of fog node $j$. We have $\lambda_j \le \mu_j$.

Compared to CDCs, each fog node can only have limited computing resources. It is generally impossible to always allow each fog node to process all the received workload. We now consider the cases that fog node $j$  processes only a portion $\alpha_j$, $0 \le \alpha_j < 1$, of its received workload 
and forwards the remaining $1-\alpha_j$ of its workload to CDCs, i.e., we still require $\alpha_j \lambda_j < \mu_j$.
We can write the expected response time for fog node $j$ as:
    \begin{eqnarray}
    {R^{W3}_{j}} \left( \alpha_{j} \right) &=& \tau^u_{j} + {\alpha_j} \left({1 \over \mu_j - \alpha_{j} \lambda_j} \right) + \left( 1 - {\alpha_{j}} \right) \tau^f.
    \label{eq_ResponseTime_W3}
    \end{eqnarray}



Consider the solution of problem (\ref{eq_OptimProb_ResponseTimeSingleNode}) by substituting the response time equation in (\ref{eq_ResponseTime_W3}).
%
%
We can observe that problem (\ref{eq_OptimProb_ResponseTimeSingleNode}) is a convex optimization problem, and hence can be solved using standard approaches. 
We omit the detailed derivation and directly present the solutions of these problems as follows. The minimum response time for users associated with fog node $j$ is $R^{W3}_j\left( \alpha^{*}_{j}\right)$, where $\alpha^{*}_{j}$ has the following closed-form solution:
\begin{eqnarray}
\alpha^{*}_{j} =\left\{ {\begin{array}{*{20}{l}}
1,\;\;\;\;\;\;\;\;\;\;\;\;\;\;\;\;\;\;\;\;\;\;\;\;\;\;\;\;\; \mbox{if } \mu_j < {\lambda_j \over \tau^f + 1}, \\
{{1\over \lambda_j} \left( w^S_j e_j \over \bar \eta_j - e_j w^D_j \right)}, \;\;\;\;\;\;\; \mbox{if } \mu_j \ge {\chi_j \over 2 \chi_j - \lambda_j \left( 1 - \tau^f \right)}, \\
{\mu_{j} \over \lambda_j} - {\mu_j \over \lambda_{j}}\sqrt{1 - {\lambda_{j} \over \mu_j} \left( 1 - \tau^f \right)}, \;\;\;\mbox{Otherwise }, 
\end{array}} \right. 
\label{eq_SolutionP1_1Fognode}
\end{eqnarray}
where $\chi_j \triangleq {w^S_j e_j \over \bar \eta_j - e_j w^D_j}$ is the maximum amount of workload that can be processed by fog node $j$ under power efficiency constraint $\eta_{j} (\alpha_j) \le \bar \eta_j$.

\begin{figure}
\centering
\includegraphics[width=1.7 in]{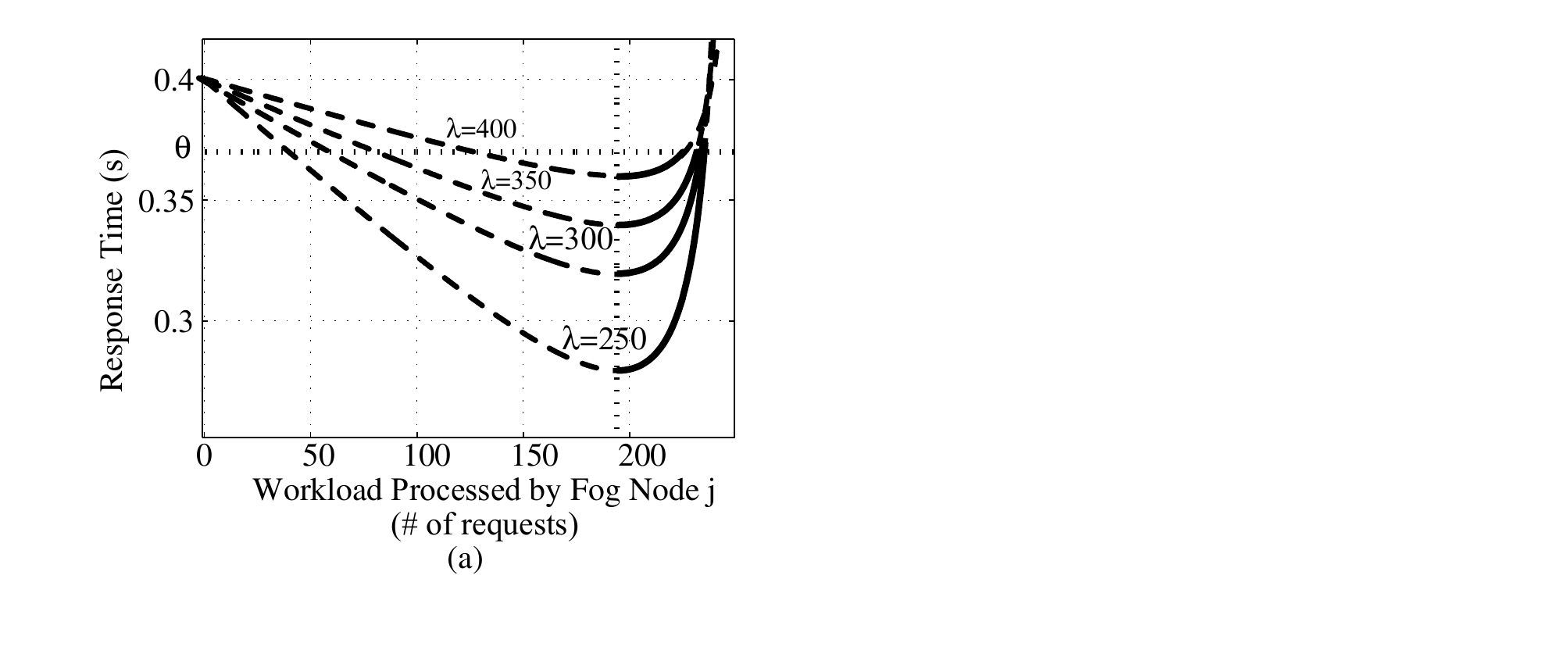}
\includegraphics[width=1.7 in]{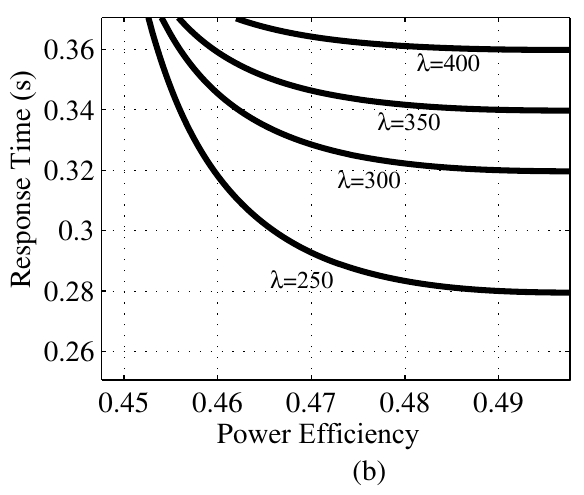}
\vspace{-0.2in}
\caption{(a) Response time under different amounts of workload processed by fog node $j$, (b) response time under different power efficiency values.}
\label{Figure_SingleNodePowerConsumptionTimeTradeoff}
\vspace{-0.2in}
\end{figure}




\subsection{Tradeoff between Response Time and Power Efficiency}
In Figure \ref{Figure_SingleNodePowerConsumptionTimeTradeoff}(a), we consider a single fog node serving 5 users, and we compare the response time under different amounts of workload (number of requests) processed by the fog node. There exists an optimal amount of workload to be processed by fog node $j$ that minimizes the response time. As observed in (\ref{eq_etaj}), the power consumption for the fog node to process one unit of workload decreases with the total amount of processed workload. In many practical applications, there is a maximum tolerable response time for users. We can therefore observe that the power efficiency maximization solution for the fog node in this case will be achieved when the response time approaches the maximum tolerable service response time $\theta$. In Figure \ref{Figure_SingleNodePowerConsumptionTimeTradeoff}(a), we use a solid line to highlight the segment between the response time minimization solution and the power efficiency maximization solution at the maximum tolerable response time $\theta$. 
We can observe 
a fundamental tradeoff between the response time and the power efficiency of the fog node. This tradeoff can be characterized by substituting (\ref{eq_ResponseTime_W3}) into (\ref{eq_OptimProb_ResponseTimeSingleNode}), as shown in Figure \ref{Figure_SingleNodePowerConsumptionTimeTradeoff}(b). We can observe that starting from the power consumption minimization point, the response time decreases with the power consumption for the fog node to process each unit (request) of workload. As the power consumption of the fog node continues to grow, the rate of reduction in the response time decreases. This means that, for non-delay-sensitive applications such as voice/image processing services (e.g., voice/image recognition), the fog node can choose a low power consumption solution as long as the resulting response time is tolerable for the users. On the other hand, for delay-sensitive applications such as online gaming and virtual reality (VR), it is ideal for the fog node to choose a high power consumption solution to satisfy users' low latency requirement. In Figure \ref{Figure_SingleNodePowerConsumptionTimeTradeoff}(b), we also present the tradeoff solutions with different workload arrival rates at fog node $j$. We can observe that the response time increases with the workload arrival rate under a given power efficiency. The higher the workload arrival rate, the smaller the changes in the response time. 
%
As the amount of workload processed by a fog node approaches its maximum processing capability, the response time approaches infinite. In other words, allowing the fog node to handle all its arriving workload cannot always reduce the response time for end-users especially when the amount of workload to be processed by the fog node cannot be carefully chosen. 

\vspace{-0.15in}
\section{Cooperative Fog Computing}
\label{Section_MultipleFognodes}

\subsection{Response Time Analysis for Cooperative Fog Computing with $N$ Fog Nodes}
\label{Subsection_CoopFogCompForwarding}
In cooperative fog computing network,
We introduce a fog node cooperation strategy, referred to as offload forwarding. In this strategy, each fog node can forward a part or all of its offloaded workload to multiple neighboring fog nodes in the fog and/or help multiple other fog nodes process their workloads.  
Each node $j$ divides its received workload into $N+1$ partitions: $\varphi_{j1}, \varphi_{j2}, \ldots, \varphi_{jN}, \varphi_{jc}$ where $\varphi_{jc}$ is the workload forwarded to the remote CDC and $\varphi_{ji}$, $i\in {\cal F}\backslash\{j\}$, is the workload forwarded to fog node $i$ (this includes $\varphi_{jj}$, the workload processed by node $j$ itself). We denote $\bvarphi_{j \bullet} \triangleq \langle \varphi_{jk} \rangle_{k \in {\cal F}}$. 
Note that it is not necessary for each fog node to always forward a non-zero workload to other fog nodes, i.e., $\varphi_{ji}=0$ means that 
fog node $i$ does not process any workload for fog node $j$. We refer to $\bvarphi_{j \bullet}$ as the request vector of fog node $j$. We also refer to $\bvarphi_{ \bullet i} = \langle \varphi_{ji} \rangle_{j \in {\cal F}}$ as the service vector of fog node $i$. Let $\bvarphi = \langle \varphi_{ji} \rangle_{i, j \in {\cal F}}$ be the workload processing matrix for the entire fog.
%
We have $0 \le \varphi_{jk} \le 1$ and $\sum\limits_{k \in {\cF}} \varphi_{jk} \le 1$, $\forall j \in \cF$. The response time of fog node $j \in \cF$ can then be written as
\begin{eqnarray}
\lefteqn{
%
R^{C}_{j} \left( \xi_j, \bphi_{j\bullet} \right)= \tau^u_{j} + 
}  \nonumber\\
&& {1 \over \sum_{i \in {\cal F}}\lambda_{i}}  \sum\limits_{i\in \cF} \phi_{ji} \left( \tau_{ji} + {1 \over  \mu_{i} - \sum\limits_{k \in \cF} \phi_{ki} } \right) 
+ \varphi_{jc} \tau^{c},
\label{eq_RCoopFogComp}
\end{eqnarray}
where $\varphi_{jc} = 1 - \sum_{i\in {\cal F}} \varphi_{ji}$, $\phi_{jk} = \lambda_{j}\varphi_{jk}$ is the amount of workload processed by fog node $k$ for fog node $j$. Please see \cite{XY2017InfocomFogComp} 
for more detailed explanation of 
 (\ref{eq_RCoopFogComp}).  
Note that if fog node $j$ cannot help other fog nodes to process their workload, but forward its own workload to other fog nodes to process, we have $\phi_{kj} = 0$ and $\phi_{ji} \neq 0$ $\forall k, i \in {\cal F}\setminus\{j\}$. 

%
%

We can rewrite the optimization problem in (\ref{eq_OptimProb_ResponseTimeMultiNode}) as follows:
\begin{eqnarray}
&& \min\limits_{\bphi_{1\bullet}, \ldots, \bphi_{N\bullet}} \sum\limits^N_{j=1} R^{C}_{j} \left( \xi_{j}, \bphi_{j\bullet} \right)
\label{eq_OptimProb_ResponseTimeMultiNodeC3}\\
&& \;\;\; \mbox{s.t.} \; \sum\limits_{k \in {\cF}} \phi_{jk} + \phi_{jc} = \lambda_j,
\label{eq_MultiNodeC3_UnsepConstraint}\\
&& \;\;\;\;\;\;\;\;\; \sum_{k \in \cal F} \phi_{kj} \le \chi_j \; \mbox{ and } \; 0\le \phi_{jk}\le \lambda_j, \forall k, j \in {\cal F}. 
\label{eq_MultiNodeC3_SepConstraint}
\end{eqnarray}

%

It can be observed that, in order for each fog node $j$ to calculate the portions of workload to be forwarded to other fog nodes, fog node $j$ needs to know the workload processing capabilities and the workload arrival rates of all the other fog nodes, which can be private information and impossible to be known by fog node $j$. In the next section, we will propose a distributed optimization framework to allow all the fog nodes to jointly optimize the average response time of the fog without disclosing their private information.   

\subsection{Response Time and Power Efficiency Tradeoff for Cooperative Fog Computing Networks}
\label{Subsection_DelayPowerTradeoffCoopFogComp}
\begin{figure}
\centering
\includegraphics[width=1.7 in]{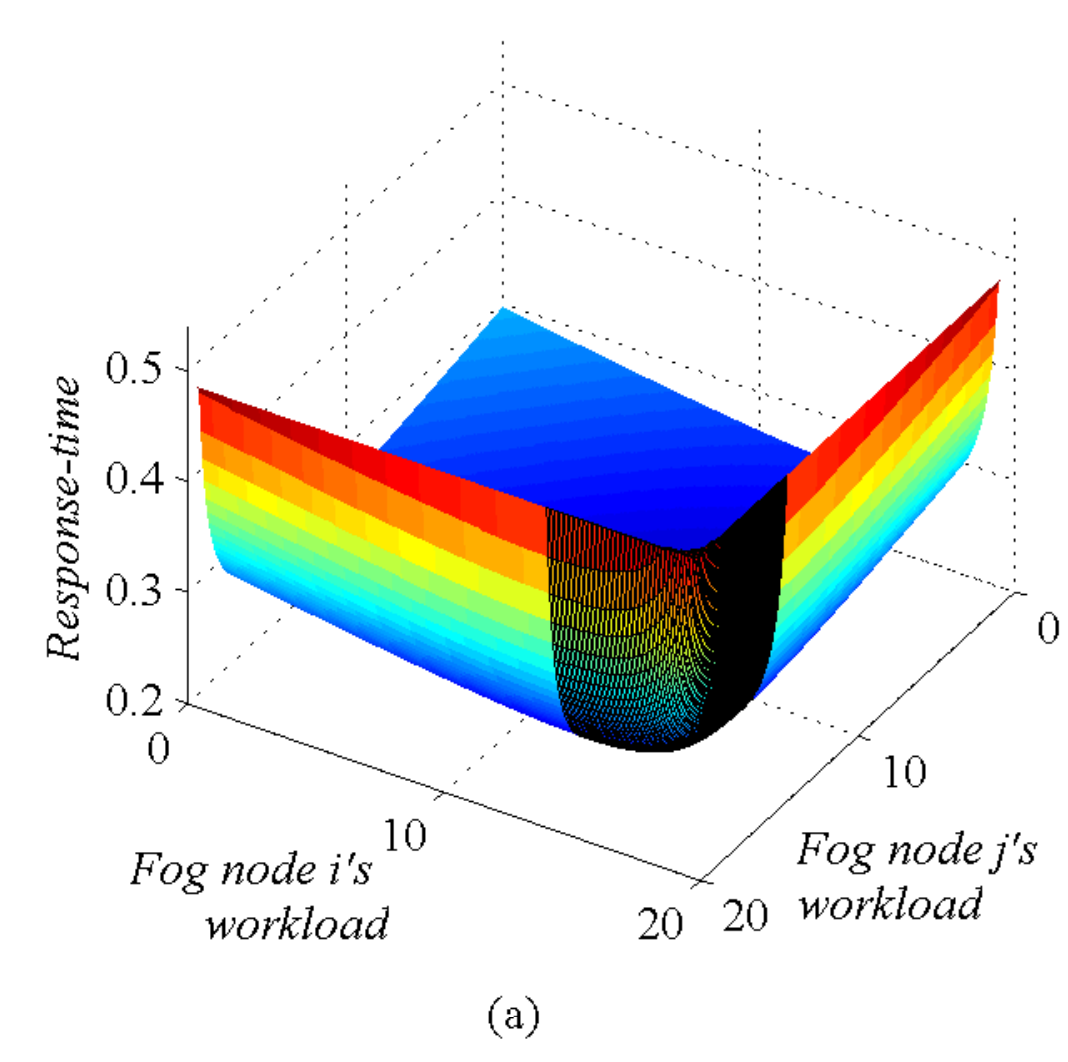}
\includegraphics[width=1.7 in]{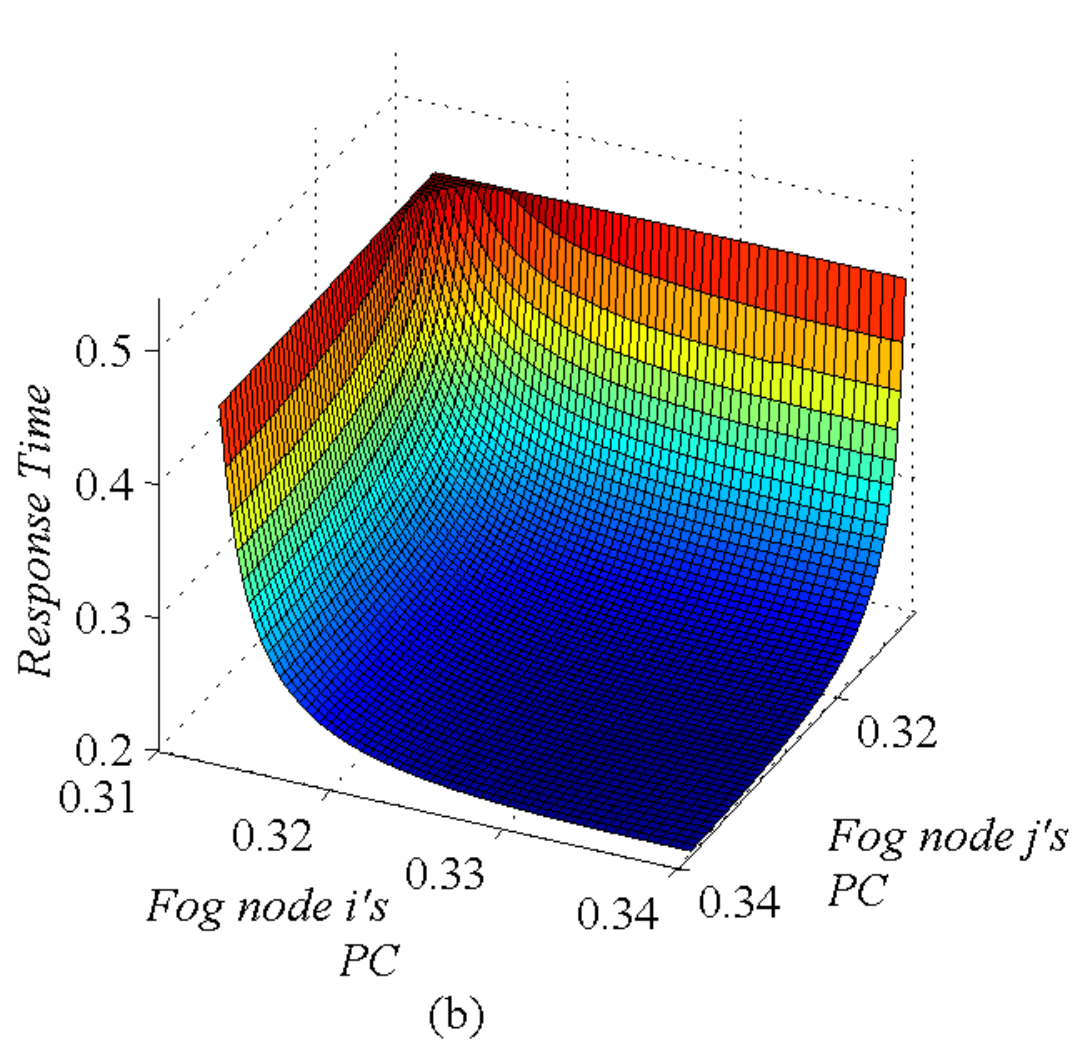}
\caption{Response time under different amounts of processed workload and power consumptions (PC) for each fog node to offload one unit of workload.}\vspace{-0.2in}
\label{Figure_MultiNodePowerConsumptionTimeTradeoff}
\end{figure}

In Figure \ref{Figure_MultiNodePowerConsumptionTimeTradeoff}(a), we present the minimum response time of the fog in a cooperative fog computing network derived from solving problem (\ref{eq_OptimProb_ResponseTimeMultiNodeC3}). Note that the workload processed by each fog node can consist of both its own received workload and the workload sent from other fog node. We can observe that the response time of the fog is closely related to the amount of workload processed by each fog node. We also use black grid to highlight the area between the response time minimization solution and the power efficiency maximization solution with a given maximum tolerable response time in Figure \ref{Figure_MultiNodePowerConsumptionTimeTradeoff}(a). By substituting the power efficiency defined in (\ref{eq_etaj}) into (\ref{eq_OptimProb_ResponseTimeMultiNodeC3}), we can also present the relationship between the fog's response time and each fog node's power efficiency for a two-node cooperative fog computing network with offload forwarding in Figure \ref{Figure_MultiNodePowerConsumptionTimeTradeoff}(b). Similar to the single-node fog computing, we can observe a fundamental tradeoff between the response time of all the users served by the fog and the power efficiency of each fog node. In addition, we can observe that by allowing offloading forwarding, even if the power consumption of each fog node to process each unit of workload has been limited to a very small value, it is still possible to achieve the response time constraint if there exist other nearby fog nodes with users that are more delay tolerant.


\section{Distributed Optimization for Cooperative Fog Computing}
\label{Section_DistributedADMM}
As mentioned previously, deciding the proper amount of workload to be processed by each fog node is essential to achieve the optimal response time and power efficiency tradeoff for fog computing networks. Unfortunately, solving problem (\ref{eq_OptimProb_ResponseTimeMultiNodeC3}) involves carefully deciding the amounts of workload processed and forwarded by every individual fog node according to global information such as the computational capacities of all the fog nodes and the round-trip workload transmission latency between any two fog nodes as well as that between fog nodes and cloud. Deploying a centralized controller to collect all these global information and calculate the optimal service and request vectors for all the fog nodes may result in a huge communication overhead and intolerably high information collection and processing delay. In addition, it can also be  observed that (\ref{eq_OptimProb_ResponseTimeMultiNodeC3}) is non-smooth and therefore cannot be solved by traditional optimization approaches that can only handle smooth objective functions. 

To address the above challenges, we need to develop a distributed framework that can solve problem (\ref{eq_OptimProb_ResponseTimeMultiNodeC3}) with the following two main design objectives:

\noindent
{\em O1) Distributed and Scalable}: we would like to develop a framework that can separate the optimization problem in (\ref{eq_OptimProb_ResponseTimeMultiNodeC3}) into $N$ sub-problems each of which can be solved by each fog node using its local information. The framework should also be scalable in the sense that the computation complexity for each fog node to solve its sub-problem should not increase significantly with the number of fog nodes that have the potential to cooperate with each other.

\noindent
{\em O2) Privacy Preserving}: Each fog node may not be willing to reveal its private proprietary information such as the maximum computational capacity and the round-trip workload transmission latency to others.

We propose a novel distributed optimization framework based on dual decomposition in which problem (\ref{eq_OptimProb_ResponseTimeMultiNodeC3}) will be first converted into its Lagrangian form and then the converted problem will be decomposed into $N$ subproblems each of which can be solved by an individual fog node using its local information. The optimization of all the subproblems will be coordinated through dual variables sent to a workload forwarding coordinator (WFC) which can be established by the cloud data centers or deployed as one of the virtualized components in the cloud.
We propose two distributed algorithms: subgradient method with dual decomposition and distributed ADMM-VS, both of which can achieve the global optimal solution of problem (\ref{eq_OptimProb_ResponseTimeMultiNodeC3}) and satisfy objectives O1) and O2).

\subsection{Subgradient Method with Dual Decomposition}
Before we introduce the algorithm, we need to first remove the inequality constraints in problem (\ref{eq_OptimProb_ResponseTimeMultiNodeC3}) by introducing a set of indicator functions. In particular, let us introduce $N+1$ indicator functions that include each of separable inequality constraints in (\ref{eq_MultiNodeC3_SepConstraint}) and incorporate these indicator functions into the objective function of problem (\ref{eq_OptimProb_ResponseTimeMultiNodeC3}). 
More specifically, we define $\cG_i = \{\bphi_{\bullet i}: \sum_{k\in {\cal F}} \phi_{ki} \le \chi_i, 0 \le \phi_{ki} \le \lambda_k, \; \forall k \in {\cal F}\}$ as the polyhedra of each constraint corresponding to fog node $i$ in problem (\ref{eq_OptimProb_ResponseTimeMultiNode}) where $\bphi_{\bullet i} = \langle \phi_{ki} \rangle_{k \in {\cal F} \backslash\{i\}}$ is the vector of amounts of workload to be processed by fog node $i$ for other fog nodes. We define an indicator function ${\bf I}_{\cG_i} \left( \bphi_{\bullet i} \right)$ 
\begin{eqnarray}
{\bf I}_{\cG_i} \left( \bphi_{\bullet i} \right) = \left\{ {\begin{array}{*{20}{c}}
{0,} & \bphi_{\bullet i} \in \cG_i, \\
{+\infty,} & \bphi_{\bullet i} \notin \cG_i.
\end{array}} \right.
\end{eqnarray}

By including the above indicator functions into the objective function of our optimization problem, we can convert the original problem (\ref{eq_OptimProb_ResponseTimeMultiNode}) with inequality constraints into the following optimization problem without inequality constraints.
\begin{eqnarray}
&& \min\limits_{\bphi_{1 \bullet}, \ldots, \bphi_{N \bullet}} \sum_{i\in {\cal F}} R^{C}_{i} \left(\xi_i, \bphi_{i \bullet}\right) + \sum_{j\in {\cal F} \cup\{c\} } {\bf I}_{\cG_j} \left( \bphi_{\bullet j} \right) 
\label{eq_OptimProb_ResponseTimeMultiNode_IndicatorFunc}\\
&& \;\;\;\; \mbox{s.t.} \;\;\; \sum_{k\in {\cal F}\cup\{c\}} \phi_{ik} = \lambda_i, \forall i \in {\cal F}. \nonumber
\end{eqnarray}

We can then write the Lagrangian form of problem (\ref{eq_OptimProb_ResponseTimeMultiNode_IndicatorFunc}) as
\begin{eqnarray}
\lefteqn{ {\cal L} \left( \bphi_{1 \bullet}, \ldots, \bphi_{N \bullet}, \bLambda \right) } \nonumber \\
&=& \sum\limits_{i\in {\cal F}} R^{C}_{i} \left(\xi_i, \bphi_{i \bullet} \right) + \sum\limits_{j\in {\cal F} \cup \{c\}} {\bf I}_{\cG_j} \nonumber \\
&& - \left(  \sum\limits_{j\in {\cal F}\cup\{c\}} \bLambda \bphi_{\bullet j} - \bLambda \blambda^\dag \right)
\label{eq_SubgradientLagrangianForm}
\end{eqnarray}
where $\bLambda$ is the vector of dual variables, $\blambda = \langle \lambda_i \rangle_{i\in {\cal F}}$ and $\cdot^\dag$ is the transpose.  

We can therefore write the optimization of the Lagrangian form as follows:
\begin{eqnarray}
\langle \bphi_{1 \bullet}, \ldots, \bphi_{N \bullet} \rangle = \arg \min_{\bphi_{\bullet j}} {\cal L} \left( \bphi_{1 \bullet}, \ldots, \bphi_{N \bullet}, \bLambda^{t} \right).
\label{eq_Subgradient_Langrangian}
\end{eqnarray}

The dual variable $\bLambda$ can be updated using
\begin{eqnarray}
\bLambda^{t+1} = \bLambda^{t} - \varrho_t \left(  \sum\limits_{j\in {\cal F}\cup\{c\}} \bLambda \bphi_{\bullet j} - \bLambda \blambda^\dag \right),
\label{eq_DualUpdating}
\end{eqnarray}
where $\varrho_t > 0$ is the step-size of the iteration.

We can observe that 
(\ref{eq_Subgradient_Langrangian}) can be reduces to solving $N$ individual sub-problems each of which can be solved by an individual fog node $i$ by optimizing the vector of workloads $\bphi_{\bullet i}$ to be processed using its local computing resources, i.e., each fog node $i$ decides values of $\bphi_{\bullet i}$ and $\varphi_{ic}$ by solving the following sub-problem:
\begin{eqnarray}
\langle\bphi^{t+1}_{\bullet i}, \varphi^{t+1}_{ic}\rangle = \arg \min_{\langle\bphi_{\bullet i}, \varphi_{ic}\rangle} {\cal L}_{S_{i}} \left( \bphi_{\bullet i}, \varphi_{ic}, \bLambda^t \right),
\end{eqnarray}
where ${\cal L}_{S_{i}} \left( \bphi_{\bullet i}, \varphi_{ic}, \bLambda^t \right) = S_{i} \left( \bphi_{\bullet i} \right) + {\bf I}_{{\cal G}_i} \left( \bphi_{\bullet i} \right) - \bLambda^t \bphi_{\bullet i} - \Lambda^t_i \varphi_{ic}$ and $S_{i} \left( \bphi_{\bullet i} \right)$ is given by
\begin{eqnarray}
S_{i} \left( \bphi_{\bullet i}, \varphi_{ic} \right) &=& \tau^u_{i} + { 1 \over \sum_{i\in {\cal F}} \lambda_{i}} \sum\limits_{j\in {\cal F}} \bphi_{ji} \left( \tau_{ji} + {1 \over u_i - \sum\limits_{k\in {\cal F}}\bphi_{ki} } \right) \nonumber \\
&&  + \varphi_{ic} \tau^{c}.
\label{eq_SeparateFogNode}
\end{eqnarray}

We can prove the following result.
\begin{theorem}
\label{Theorem_Convex}
The Lagrangian form of the objective function of our optimization problem in (\ref{eq_OptimProb_ResponseTimeMultiNodeC3}) is separable and convex.
\end{theorem}
\begin{IEEEproof}
We can directly prove the separability of the Lagrangian problem in (\ref{eq_SubgradientLagrangianForm}) by verifying ${\cal L} = \sum_{i\in {\cal F}} {\cal L}_{S_{i}}$.
Let us now prove that the objective function of problem (\ref{eq_OptimProb_ResponseTimeMultiNodeC3}) is also convex. It can be directly shown that the domain of variables in the objective function of (\ref{eq_OptimProb_ResponseTimeMultiNodeC3}) is a polyhedra which is a convex set. We can also show that the second derivative of each individual item in $S_{i} \left( \bphi_{\bullet i} \right)$ is always positive which means that it is a convex function with respect to each individual variable. Following the property that a nonnegative weighted sum of convex function $f = \sum^N_{i=1} c_i f_i$, $f: {\bf R}^N \rightarrow {\bf R}$ is convex if and only if $f_i$ is convex and $c_i$ is a constant for all $i \in \{1, 2, \ldots N\}$, we can prove that the objective function of problem (\ref{eq_OptimProb_ResponseTimeMultiNode}) is convex. This concludes the proof.
\end{IEEEproof}

It has been proved that if step-size $\varrho_t$ satisfies the following diminishing conditions $\sum^\infty_{t=1} \varrho^2_t < \infty$ and $\sum^\infty_{t=1} \varrho^2_t =\infty$,
the subgradient method with dual decomposition is guaranteed to converge to the optimal solution\cite{Falsone2017DualDecomp}. A common choice of step-size is $\varrho_t = \bar \varrho/\sqrt{k}$ where $\bar \varrho$ is a constant. 
One of the main advantage of the subgradient method is its low computational complexity for each fog node. However, it has been proved that the convergence rate of subgradient method is given by $O\left({1/\sqrt{t}}\right)$\footnote{We follows Bachmann-Landau notations: $f = O (g)$ if $\lim\limits_{n \to \infty} {f(n) \over g(n)} < +\infty$.} which is slow. 

\subsection{Distributed ADMM via Variable Splitting}
In this section, we propose a distributed optimization framework based on distributed ADMM-VS. 
Similarly, we introduce an indicator function ${\bf I}_{\cG_c} \left( \bpsi \right)$ to characterize the inseparable constraint in (\ref{eq_MultiNodeC3_UnsepConstraint}) 
\begin{eqnarray}
{\bf I}_{\cG_c} \left( \bpsi \right) = \left\{ {\begin{array}{*{20}{c}}
{0,} & \bpsi \in \cG_c, \\
{+\infty,} & \bpsi \notin \cG_c,
\end{array}} \right.
\end{eqnarray}
where $\bpsi = [\bpsi_1, \bpsi_2, \ldots, \bpsi_N]$, $\cG_c = \{\bpsi: \sum^N_{i=1} I_N \bpsi_{i} \le 1\}$, $I_N$ is an identity matrix with size $N$, $\bpsi \in {\bf R}^{N\times N}$, $\bpsi_i \in {\bf R}^{N}$.
%
%
We can show that the solution of the optimization problem in (\ref{eq_OptimProb_ResponseTimeMultiNode_IndicatorFunc}) is equivalent to solving the optimization problem with the following augmented Lagrangian form with two blocks of random variables. We can write the $\bphi$-optimization subproblem as
\begin{eqnarray}
\lefteqn{ \bphi^{t+1} = \arg \min_{\bphi} {\cal L}_{\rho} \left( \bphi_{\bullet 1}, \bphi_{\bullet 2}, \ldots, \bphi_{\bullet N}, \bpsi^t, \bLambda^t \right) }\nonumber \\
&=& \arg \min_{\bphi} \sum\limits_{i\in {\cal F}} \left\{ R^{C}_{i} \left(\xi_i, \bphi_{i \bullet} \right) + {\bf I}_{\cG_i} \left( \bphi_{\bullet i} \right) \right. \nonumber \\
&&\;\;\;\; \left. + {\rho \over 2} \| \bphi_{\bullet i} - \psi^t_i + \Lambda^t_i \|^2_2 \right\},
%
\label{eq_PhiUpdating}
\end{eqnarray}
where $\rho$ is the augmented Lagrangian parameter and $\bLambda$ is the vector of the dual variables. 

We can write the $\bpsi$-updating problem as
\begin{eqnarray}
\bpsi^{t+1} = \arg \min\limits_{\bpsi} {\rho \over 2} \| \bphi^{t+1} - \bpsi^t + \bLambda^t \|^2_2 + {\bf I}_{\cG_c} \left( \bpsi \right). 
\label{eq_PsiUpdating}
\end{eqnarray}

The dual variable update sub-problem can then be written as follows
\begin{eqnarray}
\bLambda^{t+1} = \bLambda^{t} - \rho \left( \bphi^{t+1} - \bpsi^{t+1} \right).
\label{eq_DualUpdating}
\end{eqnarray}

We can observe that the subproblem optimization in (\ref{eq_PhiUpdating})--(\ref{eq_DualUpdating}) is equivalent to the form of the traditional ADMM with two random variables: $\bphi$ and $\bpsi$.
In D-ADMM-VS,
each fog node $i$ will calculate the optimal service vector $\bphi^*_{\bullet i}$ by solving the following sub-problem:
\begin{eqnarray}
&& \bphi^{t+1}_{\bullet i} = \arg \min\limits_{\phi_{\bullet i}} {\cal L}'_{S_{i}} \left( \bphi_{\bullet i}, \bpsi^t_i, \Lambda^t_i \right) 
\label{eq_FogNodeOptmSubProb}
\end{eqnarray}


%
We present the detailed description of ADMM-VS below.
\footnotesize
\smallskip
\smallskip
\hrule
\hrule
\hrule
\hrule
\vspace{0.1cm}
{\bf Algorithm 2: Distributed ADMM-VS Algorithm}
\vspace{0.1cm}
\hrule
\hrule
\begin{itemize}
\item[] \emph{Initialization}: Each fog node $i$ chooses an initial service vector $\bphi^0_{\bullet i}$ and WFC chooses an initial dual variable $\bLambda^0$.
\begin{itemize}
\item[] WHILE {t=0, 1, \ldots}
    \begin{itemize}
        \item[i)] {\it Fog node updating}: Each fog node $i$ calculates $\bphi^{t+1}_{\bullet i}$ by solving (\ref{eq_FogNodeOptmSubProb}) and then sends the resulting $\bphi^{t+1}_{\bullet i}$ and $\lambda_k$ to the WFC,

        \item[ii)] {\it WFC Updating}: WFC calculates $\bpsi^{t+1}$ by solving $\psi$-updating problem in (\ref{eq_PsiUpdating}).

        \item[iii)] {\it Dual Variable Updating}: WFC updates dual variables $\bLambda^{t+1} = \bLambda^{k} - \rho \left( \bphi^{t+1} - \bpsi^{t+1} \right)$ and sends $\bphi^{t+1}_i$ and $\Lambda^{t+1}_{i}$ to fog node $i$.
    \end{itemize}
\item[] ENDWHILE

\end{itemize}

\end{itemize}

\hrule
\hrule
\hrule
\hrule
\smallskip
\smallskip
\normalsize
%
%
We have the following result. 
\begin{theorem}
Our proposed D-ADMM-VS algorithm converges to the global optimal solution of Problem (\ref{eq_OptimProb_ResponseTimeMultiNode}) with convergence rate of $O\left(1/t\right)$.
\end{theorem}
\begin{IEEEproof}
The convergence of Algorithm 2 follows directly from the standard ADMM approach\cite{Boyd2011ADMM}. We omit the detailed description due to limit of space.
\end{IEEEproof}

\begin{figure}%
\begin{minipage}[t]{0.45\linewidth}
\centering
\includegraphics[width=1.4 in]{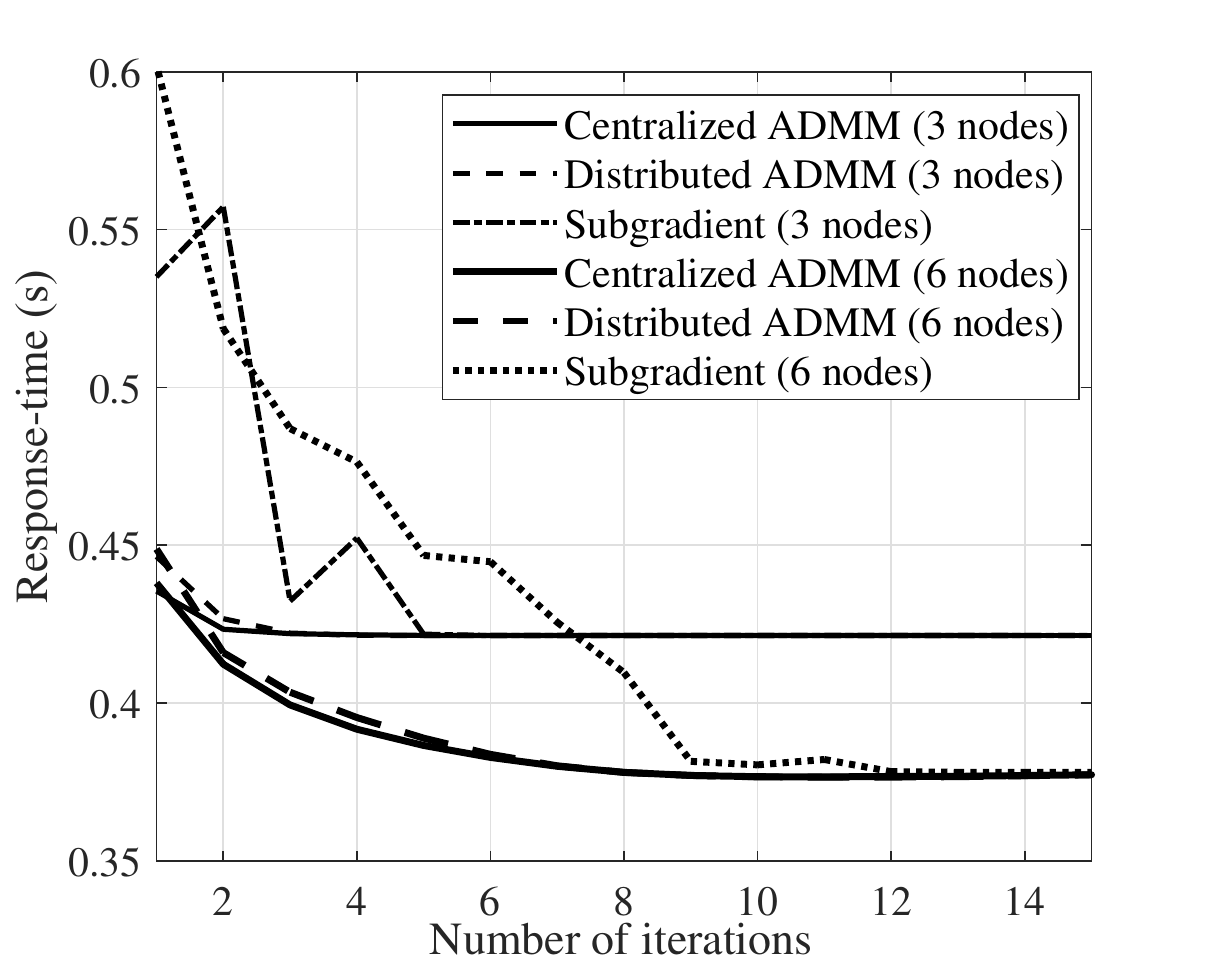}
\vspace{-0.1in}
\caption{Convergence rate of the proposed algorithms compared to a centralized ADMM approach.} 
\label{Figure_ConvergenceRate}
\end{minipage}
\begin{minipage}[t]{0.45\linewidth}
\centering
\includegraphics[width=1.6 in]{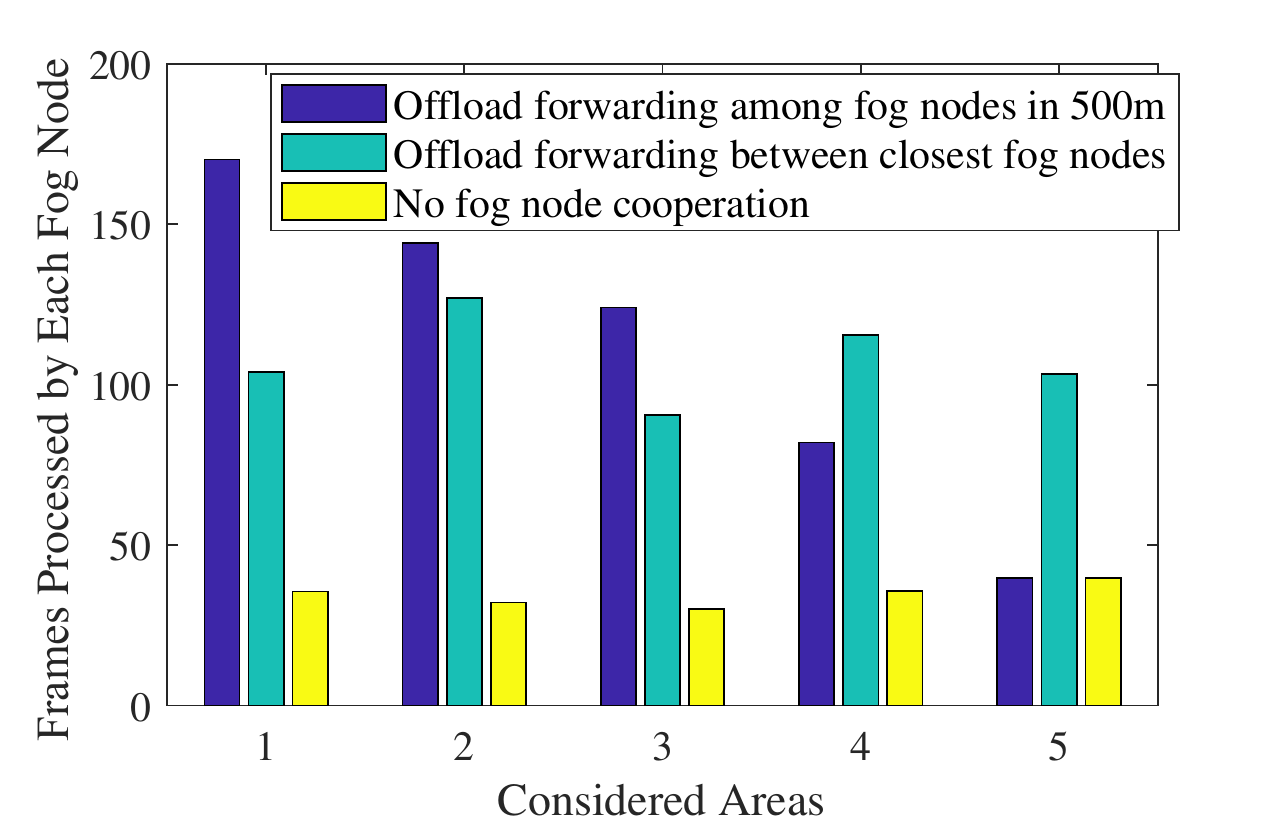}
\vspace{-0.3in}
\caption{Average workload (number of requests) processed by each fog node at different areas of consideration.}
\label{Figure_OffloadVSAreas}
\end{minipage}
\vspace{-0.2in}
\end{figure}

We evaluate the the convergence performance of our proposed algorithms in Figure \ref{Figure_ConvergenceRate}. We can observe that both our proposed algorithms can converge to the global optimal solution within the first few iterations (less than 14 iterations in both cases) as shown in Figure \ref{Figure_ConvergenceRate}. We also present the convergence rate when a centralized ADMM method in \cite{Boyd2011ADMM} is applied to solve optimization problem (\ref{eq_OptimProb_ResponseTimeMultiNodeC3}). In this method, a centralized controller can collect all the information from fog nodes and calculate the amount of workload to be processed by each fog node. We can observe that our proposed distributed ADMM-VS presents a similar convergence performance as the centralized ADMM approach and can approach the global optimal solution within first 10 iterations which is much faster than the subgradient method. 

\begin{figure}
\centering
\includegraphics[width=3.6 in]{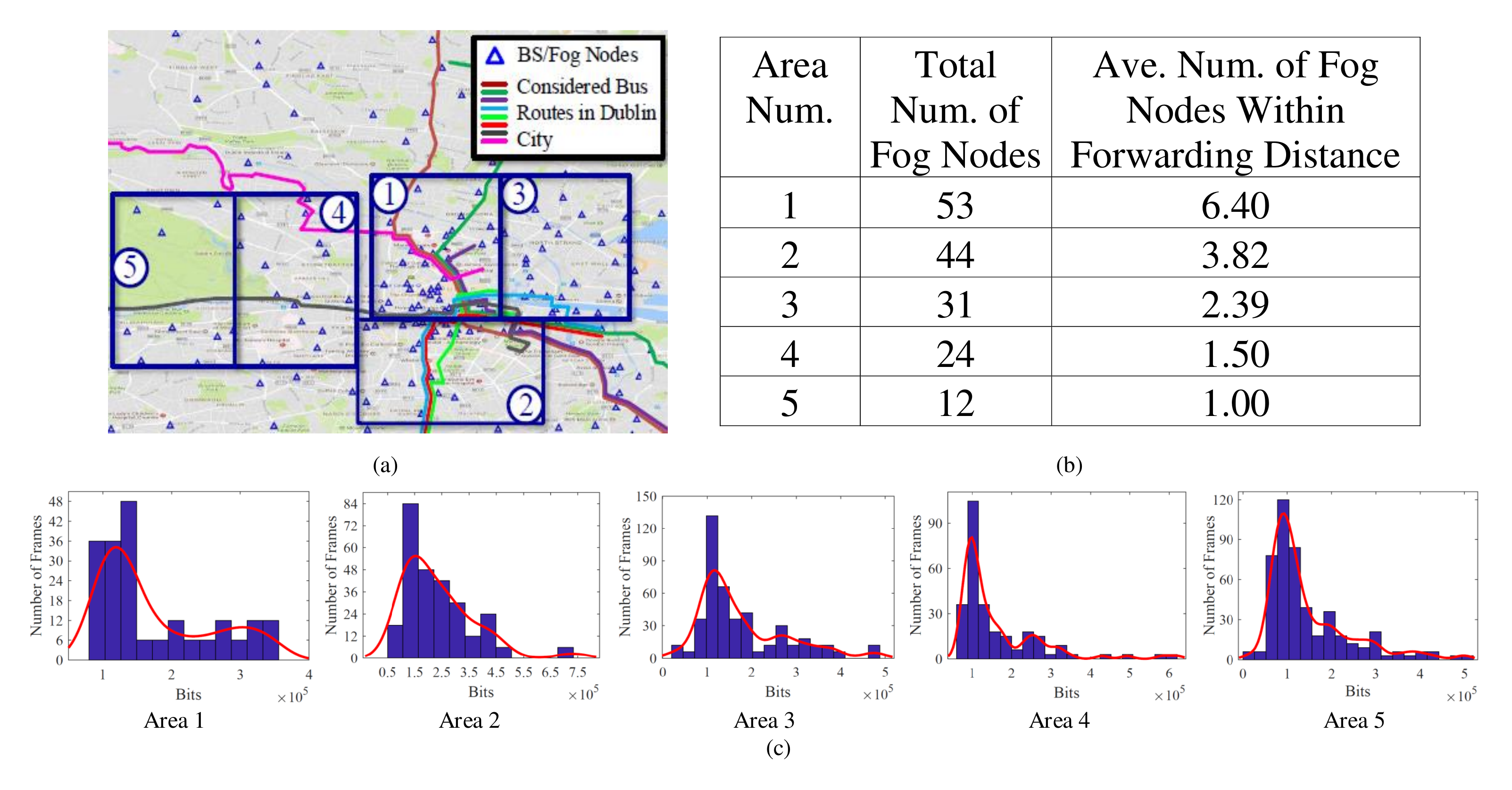}
\vspace{-0.4 in}
\caption{(a) Distribution of fog nodes, bus routes, and considered areas, (b) deployment density of fog nodes in each considered area, and (c) empirical probability distribution of traffics generated by self-driving buses in each considered area.}
\label{Figure_BSdistribution}
\vspace{-0.2in}
\end{figure}

\begin{figure}
%
\begin{minipage}[t]{0.45\linewidth}
\centering
\includegraphics[width=1.5 in]{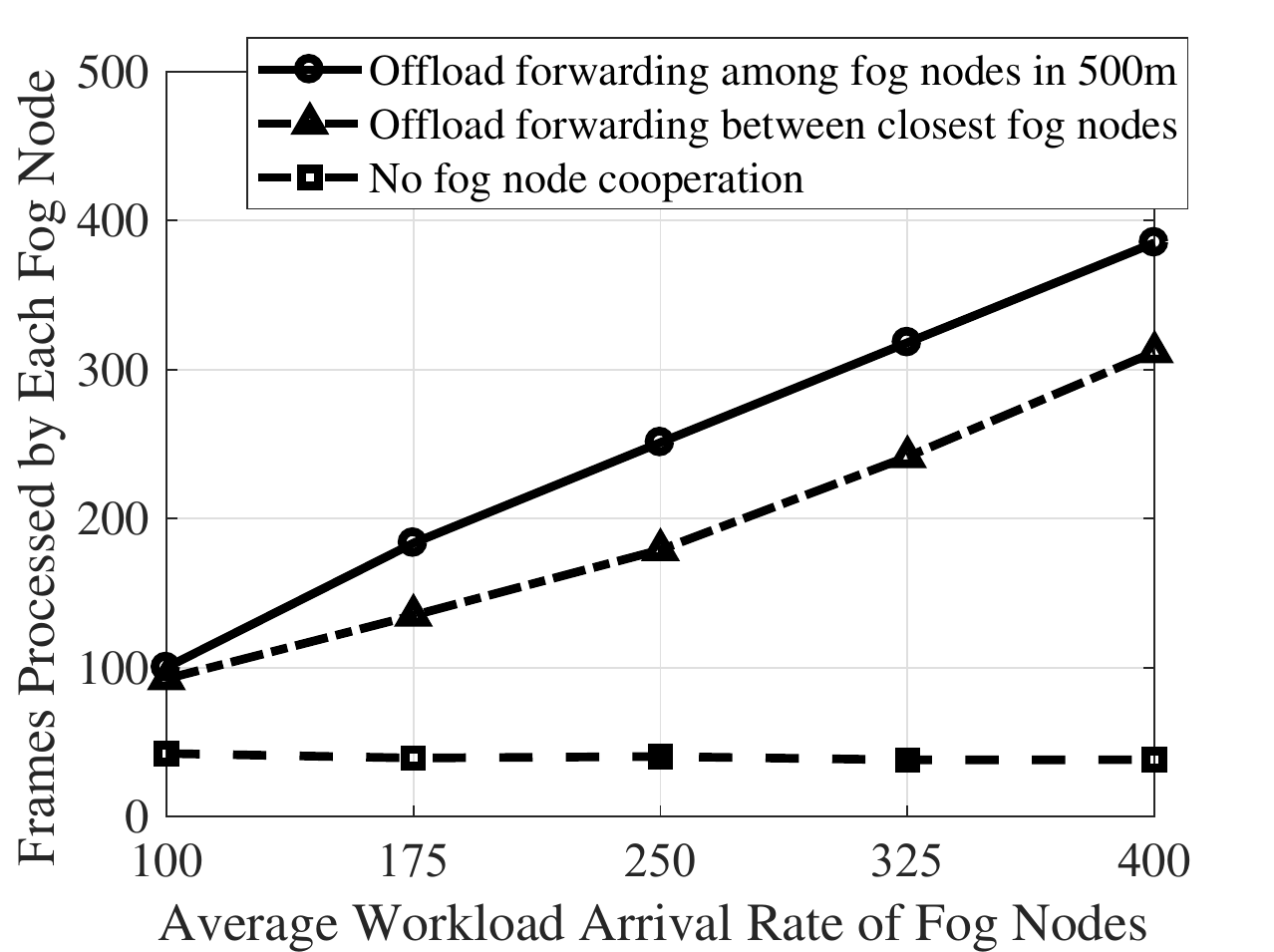}
\caption{Average workload (number of requests) processed by each fog node with different workload arrival rates.} 
\label{Figure_OffloadVSArrivalRates}
\end{minipage}
\begin{minipage}[t]{0.45\linewidth}
\centering
\includegraphics[width=1.7 in]{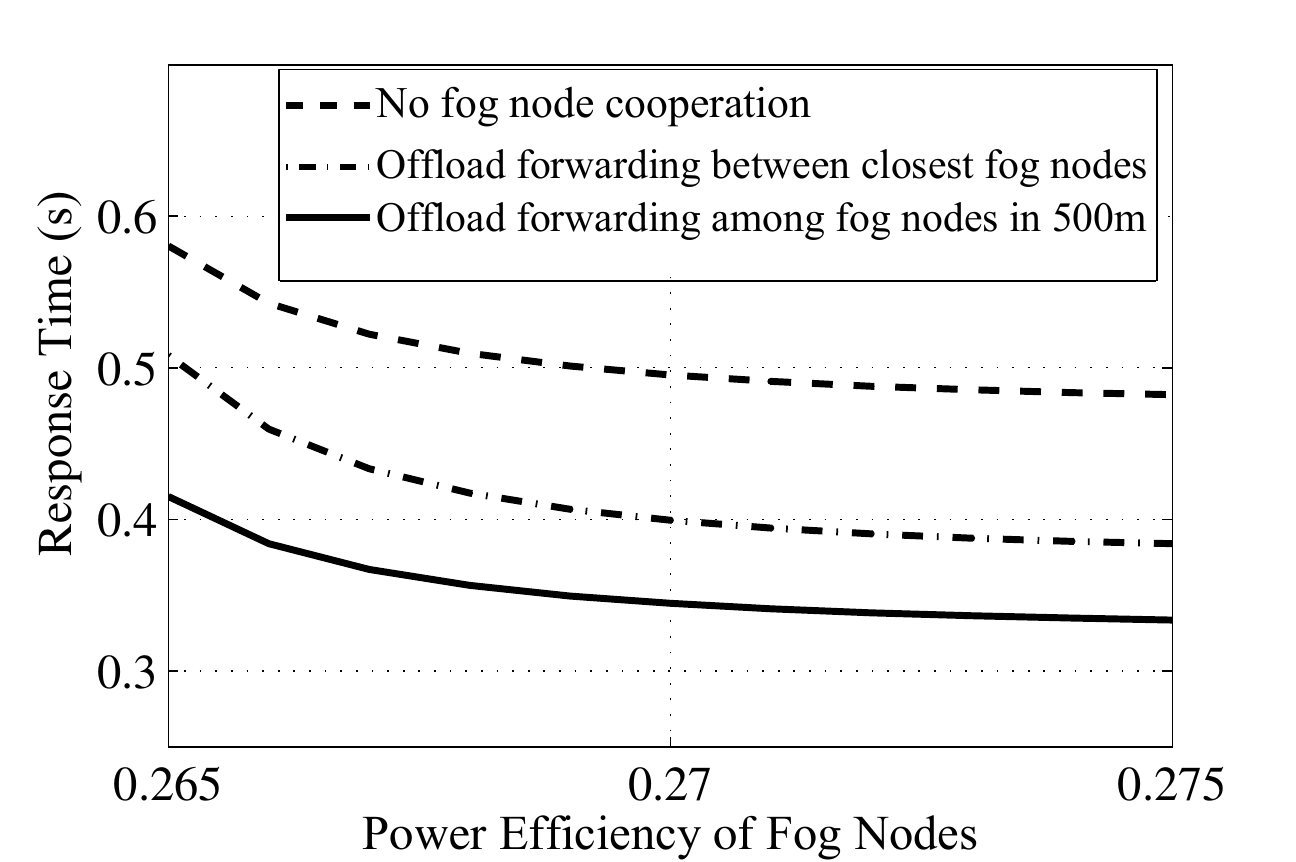}
\caption{Average response time of users with different power efficiency.}
\label{Figure_ResponseTimeVSPowerEfficiency}
\end{minipage}
\vspace{-0.3in}
\end{figure}

\section{Case Study: a City-wide Deployment of Fog Computing-supported Self-driving Bus System}
\label{Section_CaseStudy}
\subsection{Simulation Setup for Traffics Generated by Self-driving Bus}

In this section, we consider a possible implementation of a fog computing-supported self-driving bus system in the city of Dublin as a case study. A self-driving vehicle relies on a combination of sensors including cameras, sonars, radars, light detection and ranging systems (LiDARs), etc., to sense the surrounds and decide the driving behaviors. More specifically, image and sensing data collected by the sensors will be processed by a computer or processor to extract useful information such as the types of objects as well as their specific semantics that may affect the driving  decisions in various scenarios. For example, 
an autonomous vehicle must be able to detect and recognize various types of unintelligent objects such as traffic/road work signs and traffic lights as well as intelligent objects including surrounding vehicles, animals, and pedestrians. 
It is known that accurate and low-latency object recognition require significant computing power and energy supply\cite{Teichman2011SelfDrivingCar}.   
How to develop effective object recognition methods for autonomous driving vehicles is out of the scope of this paper.

In this section, we focus on the scenario that each self-driving bus relies on a fog node in proximity to process the traffic image and feedback the driving decision. We focus on the workload transmission and forwarding between vehicles and the fog nodes. It is known that, for each self-driving vehicle, the amount of data that needs to be collected and processed is different when it drives into different areas. For example, the traffic condition in the city center will be much more complex than that in the countryside. To take into consideration of the geographical diversity of the traffic data generated by each bus, 
%
we analyze the statistic feature of the traffics generated by buses operated at 8 existing routes in the city of Dublin. We generate driving videos from over 2500 stitched high-resolution street view images in the considered bus routes extracted from Google street view\cite{GoogleStreetV}. We then apply H.265 coding to compress the generated driving videos and keep track of the recorded frame rates of the compressed video to simulate the possible traffic data streaming of each self-driving bus. H.265 is a high efficiency video coding technique that can remove both temporal and spatial redundancies between adjacent frames using the enhanced hybrid spatial-temporal prediction model\cite{Sullivan2018H265}. In other words, the frame rates generated by H.265 can reflect different levels of traffic complexity as well as the impact of the driving speed and surrounding traffics at different locations.
We fit the recorded frame rates of each bus route using a Kernel-based probability distribution function. The bus driving routes, considered areas, and fog node distribution are shown in Figure \ref{Figure_BSdistribution}(a). The deployment densities of fog nodes are listed in Figure Figure \ref{Figure_BSdistribution}(b). The recorded frame sizes and fitted probability distributions of traffic generated by the self-driving bus when driving in different considered areas are presented in Figure \ref{Figure_BSdistribution}(c).



%
\subsection{Simulation Setup for a Fog Computing Network}
%
We simulate a possible implementation of fog nodes, e.g., mini-computing servers, over 200 BSs (including GSM and UMTS BSs) deployed by a primary telecom operator in the city of Dublin. 
The actual distribution and the deployment density of fog nodes are shown in Figure \ref{Figure_BSdistribution}. 
Each bus always submits the traffic images (frame-by-frame) taken by its on-board camera to its closest fog node.  In this case, the fog node installed at each BS will be responsible to receive and process the images sent by each bus and feedback the driving decision to each bus when the processing (e.g., objective recognition, tracking, and prediction) is finished. We assume each fog node can process at most 400 frames at the same time and the maximum tolerable response time of each bus is 500 ms. 
We consider two scenarios of offloading forwarding. In the first one, each fog node can only forward its workload to its closest fog node. In the second scenario, each fog node can forward part of its received workload to other fog nodes within a 500-meter range. 
We assume there exist local communication links among fog nodes and the round trip workload forwarding time between any two fog nodes within forwarding distance is the same given by $\tau_{ij} = 20ms$.

\subsection{Numerical Results}
To evaluate the performance improvement that can be achieved by allowing offload forwarding among fog nodes, we first compare the number of frames that can be processed by each fog node in the five areas highlighted in Figure \ref{Figure_BSdistribution}(a). We can observe in Figure \ref{Figure_OffloadVSAreas} that by allowing each fog node to cooperate with all the other fog nodes within a 500-meter range can significantly improve the numbers of frames processed by fog nodes. We can also observe that even when each fog node can only cooperate with its closest fog node, the average number of frames processed by each fog node can be almost doubled compared to the case without cooperation among fog nodes. Note that in Figure \ref{Figure_OffloadVSAreas}, we can also observe that in areas 4 and 5, allowing fog nodes within a 500-meter range cannot achieve a higher workload offloading performance than only allowing each fog node to cooperate with its closest neighboring fog node. This is because in both of these two considered rural areas, some fog nodes cannot have any other fog node located within the 500-meter range.

In Figures \ref{Figure_OffloadVSArrivalRates}, we consider the average workload processing capability of all 5 considered areas. We investigate the impact of fog nodes' workload arrival rates on the total amount of workload to be offloaded by the fog computing network. 
We can observe that the average number of frames that can be offloaded by each fog node increases almost linearly when the workload arrival rate is small. However, with the workload arrival rate continuing to grow, the total amount of offloaded workload that can be offloaded by fog computing network approaches to a fixed value limited by the maximum response time that can be tolerated by end-users.

In Figure \ref{Figure_ResponseTimeVSPowerEfficiency}, we present the average response time and power efficiency tradeoff curves with and without offload forwarding. 
We observe that our proposed offload forwarding significantly reduces the response time of end-users especially when the power efficiency constraint of fog node $j$ is low. With the increase of the power efficiency of fog nodes, the response time that can be provided by the fog nodes approaches a fixed value limited by the maximum workload processing capability of the fog.
\vspace{-0.2in}
\section{Conclusion}
\label{Section_Conclusion}

In this paper, the workload offloading problem was studied for fog computing networks. We investigated the relationship between users' response time and fog nodes' power efficiency. The tradeoff between these two metrics was discussed for fog computing network with and without fog node cooperation. 
For cooperative fog computing networks, we introduced a novel fog node cooperation strategy called offload forwarding. In this strategy, each fog node can forward a part of its workload to other neighboring fog nodes to further reduce the response time. 
We quantify the response time and power efficiency tradeoff for cooperative fog computing with offload forwarding. 
A distributed optimization framework based on dual decomposition has been proposed. We developed two distributed algorithms under the proposed framework. The first one is based on the subgradient method with dual decomposition and the other algorithm is based on the distributed ADMM-VS. We proved that both proposed algorithms can approach the globally optimal workload allocation solution. Finally, we have considered the possible implementation of a city-wide self-driving bus system supported by a fog computing network as a case study to verify the performance of our proposed approach.
Numerical results have shown that our proposed framework can significantly improve the performance of fog computing networks.
%
%

\section*{Acknowledgment}
Yong Xiao would like to thank Professor Luiz A. DaSilva and Dr. Jacek Kibilda at CONNECT, Trinity College Dublin to provide the BS location data of Dublin city.

\vspace{-0.1in}

\bibliography{reference}
\bibliographystyle{IEEEtran}

\begin{IEEEbiography}{Yong Xiao}(S'09-M'13-SM'15) received his B.S. degree in electrical engineering from China University of Geosciences, Wuhan, China in 2002, M.Sc. degree in telecommunication from Hong Kong University of Science and Technology in 2006, and his Ph. D degree in electrical and electronic engineering from Nanyang Technological University, Singapore in 2012. He is now a professor in the School of Electronic Information and Communications at the Huazhong University of Science and Technology (HUST), Wuhan, China. 
His research interests include machine learning, game theory, distributed optimization, and their applications in cloud/fog/mobile edge computing, green communication systems, wireless communication networks, and Internet-of-Things (IoT).
%
\end{IEEEbiography}

\vskip -2\baselineskip plus -1fil


\begin{IEEEbiography}{Marwan Krunz}(S'93-M'95-SM'04-F'10) is the Kenneth VonBehren Endowed Professor in the Department of Electrical and Computer Engineering at the University of Arizona (UA). He is also the director of the Broadband Wireless Access and Applications Center (BWAC), a joint NSF/industry consortium that includes UA (lead site), Virginia Tech, University of Notre Dame, University of Mississippi, Auburn University, and Catholic University of America, as well as 20+ members from industry and national labs. Dr. Krunz received the Ph.D. degree in electrical engineering from Michigan State University in July 1995. He joined the University of Arizona in January 1997, after a brief postdoctoral stint at the University of Maryland, College Park. In 2010, he was a Visiting Chair of Excellence (``Catedra de Excelencia") at the University of Carlos III de Madrid (Spain). He held numerous other short-term research positions at the University Technology Sydney, Australia (2016), University of Paris V, INRIA-Sophia Antipolis, France , University of Paris VI, HP Labs, Palo Alto, and US West Advanced Technologies. He is the Editor-in-Chief for the IEEE Transactions on Mobile Computing, and he previously served on the editorial boards of numerous journals, including IEEE/ACM Trans. on Networking, IEEE Trans. on Cognitive Communications and Networking, and others.
\end{IEEEbiography}

\end{document}